\documentclass[10pt,preprint]{aastex}

\begin{document}
\title{The Birthplaces of Gamma-Ray Bursts}

\author{Patrick A. Young\altaffilmark{1,2} \& Chris L. Fryer\altaffilmark{3,4}}

\altaffiltext{1}{Astronomy Department, The University of Arizona,
Tucson, AZ 85721} 
\altaffiltext{2}{X Division, Los Alamos National Laboratory, 
Los Alamos, NM 87545}
\altaffiltext{3}{Department of Physics, The University of Arizona,
Tucson, AZ 85721} 
\altaffiltext{4}{CCS Division, Los Alamos National Laboratory, 
Los Alamos, NM 87545}

\begin{abstract}
We use population synthesis to construct distributions of gamma-ray
bursts (GRBs) for different proposed progenitor models. We use a
description of star formation that takes into account the evolution of
metallicity with redshift and galaxy mass, the evolution of galaxy
mass with redshift, and the star formation rate with galaxy mass and
redshift. We compare predicted distributions with redshift and
metallicity to observations of GRB host galaxies and find that the
the simple models cannot produce the observed distributions, but 
that current theoretical models can reproduce the observations 
within some constraints on the fraction of fallback black holes 
that produce GRBs.
\end{abstract}


\section{Introduction}

With the advent of the SWIFT satellite the number of gamma-ray bursts
(GRBs) with multiwavelength observations of afterglows and host
galaxies is growing rapidly.  As the data increases, it has the
potential to allow us to probe deeper into the question of the
progenitors of these outbursts.  On the surface, much of the data seems
contradictory (for a review, see \citet{Fry07}): recent data has
confirmed that the explosions behind long-duration bursts produce
supernovae, but we now know that some don't, or if they do, they
produce dim supernovae~\citep{Fyn06,Ber06}; GRBs are associated with
star forming regions, but they appear to occur in lower mass galaxies
than their supernova counterparts~\citep{Fru06}; and although many
bursts occur at metallicities below 1/10th solar, the average
metallicity of bursts may well be closer to 1/3 to 1/2 solar
\citep{pro07}.

All of this data may not be as contradictory as it seems.  For
instance, the idea that many bursts occur in low metallicity
environments need not mean that bursts are exclusively produced by
low-metallicity stars.  If we consider that star formation is biased
towards higher redshift and lower mass galaxies than a typical $L*$
galaxy today, both of which imply lower metallicities, we might expect
such an observational bias even if there is {\it no} metallicity
dependence for GRB progenitors at all.  As our sample of well-studied 
bursts grows, it will become easier to extract reality 
from observational bias.  

Theory has long made qualitative predictions about these bursts.
Bursts are believed to come from black hole forming stars and the
stars that form black holes tend to be the most massive stars.  Thus,
it is not surprising that GRBs and supernovae do not occur in exactly
the same conditions.  Since the strong stellar winds seen at high
metallicites will prevent stars from collapsing directly to black
holes, it has long been suggested the rate of long duration bursts
will be higher at lower metallicities \citep{FWH99,Fry07}.  Because
the $^{56}$Ni yield of a GRB explosion depends sensitively on whether
the black hole is formed directly or by fallback in a weak initial
explosion, theorists have also known that bursts would not necessarily
always be accompanied by bright supernova outbursts
\citep{FWH99,FYH06}.  But to actually extract information about GRB
progenitors with the current data, we must develop a more quantitative
theory.

In this paper we perform population synthesis calculations to determine
if the common progenitor scenarios for long duration GRBs can be
distinguished based upon the observational data.
We combine the star formation rate as a function of redshift,
evolution of cosmic metal density with redshift, the mass metallicity
relation for galxies, and the specific star formation rate versus
galaxy mass.  Using a Monte Carlo technique, we sample these 
distributions to produce the expected population of gamma-ray bursts and 
supernovae as a function of metallicity, redshift, and even host 
galaxy mass.  Depending upon the progenitor, these distributions 
differ and we can use observations of bursts to constrain the 
progenitor model.

Section~\ref{calc} describes the population synthesis Monte Carlo code
used to create the simulated distributions, the descriptions of star
formation and host galaxy properties used, and the GRB progenitor
models. Section~\ref{results} compares our simulated distributions to
observed GRB rates. Section~\ref{conclusion} discusses the
implications for the nature of GRB progenitors and our ability to
constrain theoretical models.

\section{Calculations \label{calc}}

\subsection{Initial Conditions}

We calculate the observed population of gamma-ray bursts by
introducing a Monte-Carlo code that pulls from a sample of stars based
on what we know of star formation.  The fate of a massive star depends
upon its metallicity, its initial mass, and its initial spin.  To
calculate the distribution of gamma-ray bursts, we first must assume a
distribution for each of these quantities. 


Since the average metallicity of the universe decreases with
increasing redshift, the metallicity of our entire distribution of
GRBs clearly depends upon the star formation rate as a function of
redshift. The mass-metallicity relation for galaxies also requires
that we take into account the mass distribution of galaxies with
redshift and the star formation rate as a function of galaxy mass.

For our simple model, we use the functional form of the star formation
rate per unit volume as a function of redshift ($z$) from Watanabe et
al.(1999)
\begin{equation}
log [\eta(z)] = \left\{ \begin{array}{ll} 
                        A log (1+z) & {\rm for}\; z \leq z_p \\
                        A log (1+z_p) - B(z-z_p) & {\rm for}\; z>z_p
\end{array} \right.
\label{eq:sfh}  
\end{equation}
$A, B, z_p$ are parameters that describe the star formation rate.  The
star formation rises from the initial burst of star formation and
peaks at $z_p$ and then drops precipitously as we approach redshift of
0.  Our standard model for the star formation history is based on a
fit of this simple model to the current observations (including the
metallicity evolution we use based on the work of Hopkins \& Beacom
2006 - see below): ($A,B,z_p$) = (1.0,0.5,2.5).  The star formation
history remains uncertain, but its peak is now believed to lie
somewhere between 2 and 3.  We run 2 additional models probing the
dependence of our results on this history: ($A,B,z_p$) = (1.25,0.5,2.0) and
(1.0,0.1,2.5).  To get the rate of star formation in a given redshift
bin, we must also correct for the volume filling factor:
\begin{equation}
\Delta V (z) = \Delta z (1.0-\sqrt{1.0+z})^2 \times (1.0+z)^{-3/2}.
\end{equation}

For the mass-metallicity relation for galaxies we use the prescription
derived by \citet{christy04} from the Sloan Digital Sky Survey (SDSS)
galaxy sample, which has the functional form

\begin{equation}
12+\log{\rm (O/H)} = -1.492+1.847\log M_* - 0.08026(\log M_*)^2
\end{equation}

\citet{erb06} show that the shape of the mass-metallicity relation
remains essentially the same out to redshifts of $z\sim 2$, varying
primarily in the normalization of the total metal content. We assume
that the total metallicity scales along with the $\log$(O/H)
ratio. This is of course not strictly true, but it is the best
observational approximation available. We assume that this functional
form is valid to higher redshifts and shift the normalization of the
distribution according to the total metal density of the universe as a
function of redshift. For the total metal density we adopt the form in
\citet{hb06}.

\begin{eqnarray}
\dot{\rho}_* & = & (a+bz)/[1+(z/c)^d] h_{70}M_{\odot}yr^{-1}Mpc^{-3} \\
\dot{\rho}_* & = & 63.7\dot{\rho}_z \\
\rho_z(z) & = & \int^z \dot{\rho}_zdz 
\end{eqnarray}

where $\dot{\rho}_*$ is the rate of change of the stellar mass density
with redshift, $\dot{\rho}_z$ is the rate of change of the metal
density, and the coefficients $a=0.017,b=0.13,c=3.3$ and $d=5.3$ are
defined in \citet{cole01}.

We use the galaxy mass function for $z \leq 4$ from the GOODS-MUSIC
sample \citep{font06}. This is fit with a Shechter function with
redshift dependent $M*$, and $\alpha *$ described as decribed in
\citet{font06}:
\begin{equation}
f(M_{\rm galaxy}) \propto log[10^{M-M^*(z)}]^{\alpha^*(z)} e^-{10^{M-M^*(z)}}
\end{equation}
where $M$ is the log of the galaxy mass, $M^*(z) = 11.16 + 0.17z -0.07z^2$, 
and $\alpha^*(z) = -.18-0.082z$.  Above a redshift of 4, we set the mass 
function to the value at a redshift of 4.

The most difficult part of the problem is a description for the star
formation rate as a function of galaxy mass at different
redshifts. The star formation in the local universe is dominated by
dwarf galaxies, but by redshifts of 0.7 large spirals account for more
than half. We produce a rough estimate of the specific star formation
rate ($\dot{M}_{\odot}/M_*$) with respect to galaxy mass from Figure
13 of \citet{casey06}. We fit a line to the points in the figures and
take this as the specific star formation rate in three redshift bins
centered on the redshifts of the three panels of their Figure
14:
\begin{eqnarray}
log (SFR/M_{\rm galaxy}) & \propto & 1.035 log M_{\rm galaxy} \,{\rm if} z < 0.5 \\
                         & \propto & 0.8947 log M_{\rm galaxy} \,{\rm if} 0.5 < z < 0.7 \\
                         & \propto & 0.44 log M_{\rm galaxy} \,{\rm if} 1.5 < z 
\end{eqnarray}
where $M_{\rm galaxy}$ is the galaxy mass. These approximations are
crude, but at least give us a zeroth order correction to the
metallicity distribution of GRBs due to the mass-metallicity relation.

For our simple stellar mass distribution, we assume a single power 
law on the mass (M) of the star:
\begin{equation}
f(M) = M^{-\Gamma} dM, 
\end{equation}
where $\Gamma$ is set to 2.7 for our baseline case.

At this point, the uncertainties in our understanding of the evolution
of angular momentum is too great for us to make any calculations of
this quantity.  For the time being, we assume that there is always
sufficient angular momentum to make a GRB.  This provides an upper
limit on the GRB rate.

\subsection{Fates of Systems}

With the initial conditions outlined above, we produce a population of
stars with given metallicities.  To determine which is a GRB, we must
then convolve this result with our knowledge of stellar evolution.
Unfortunately, the final state of a star differs for different stellar
evolution codes.  The primary physical uncertainties in these codes
include our understanding of convection in the stars (especially when
it is coupled to nuclear burning), pulsational instabilities, and mass
loss.  Understanding rotation adds a new layer of uncertainties and we
will not study this further in the paper.  Binary stellar models are
few and far between, generally focusing on specific objects.  We will
only make rough estimates of binary progenitors in this paper.

As a test case, we first introduce a simple model that assumes 
all stars above a certain mass limit produce GRBs.  This ignores the 
fact that mass loss is required to remove the hydrogen envelope, and 
too much mass loss prevents the formation of a black hole.  We will 
use this simple model as a gauge to test our more realistic progenitor 
scenarios:  basic single star progenitors, binary star progenitors, and the 
more recent single-star models suggestied by \citet{YL05}.  

For our basic single-star models, we take the Heger et al. (2003)
paper using the strict definition of the plots where the metallicity
is given as a log plot with ``about solar'' taken for exactly solar
and the lower part of the plot taken for 10$^{-6}$ solar.  Using 
these definitions, we can predict the fates of a star as a function of
metallicity and mass.  Recall that our minimum requirements for a GRB
are that the star must (i) collapse down to form a black hole and (ii) 
lose its hydrogen envelope\footnote{We assume that the star has
sufficient angular momentum to form a GRB in all cases and hence, all
our GRB rates are upper limits.  We discuss the repurcussions of this
assumption more fully in the conclusions.}.

Our estimate of the effects of binaries is to assume that binary
systems always remove the hydrogen envelope of the collapsing star.
So our constraint for making GRBs is simply that the star must
collapse to a black hole.  Here we still use the Heger et al.(2003)
distribution, but we can now extend the formation to include those 
stars that would normally still retain a hydrogen envelope.  Note 
that specific binary scenarios may well produce distributions 
that are very different than the one used in our simple choice, 
but this provides us with a first guess of the distribution.

We also include the more recent model suggested by Yoon \& Langer 
(2005) that found that high rotation in stars could lead to 
extensive mixing that allows the near-complete burning of the 
hydrogen envelope.  We base our models on the predictions in 
their most recent results\citep{Yoo06}.

Finally, remember that we can form black holes in 2 separate ways: a
direct collapse to a black hole or the delayed collapse to a black
hole via fallback after a weak supernova explosion.  Some fraction of
weak supernova explosions will form very little $^{56}$Ni (Fryer 
et al. 2006) and we will differentiate between these two black 
hole formation mechanisms.  

\section{Results \label{results}}

\begin{figure}
\plotone{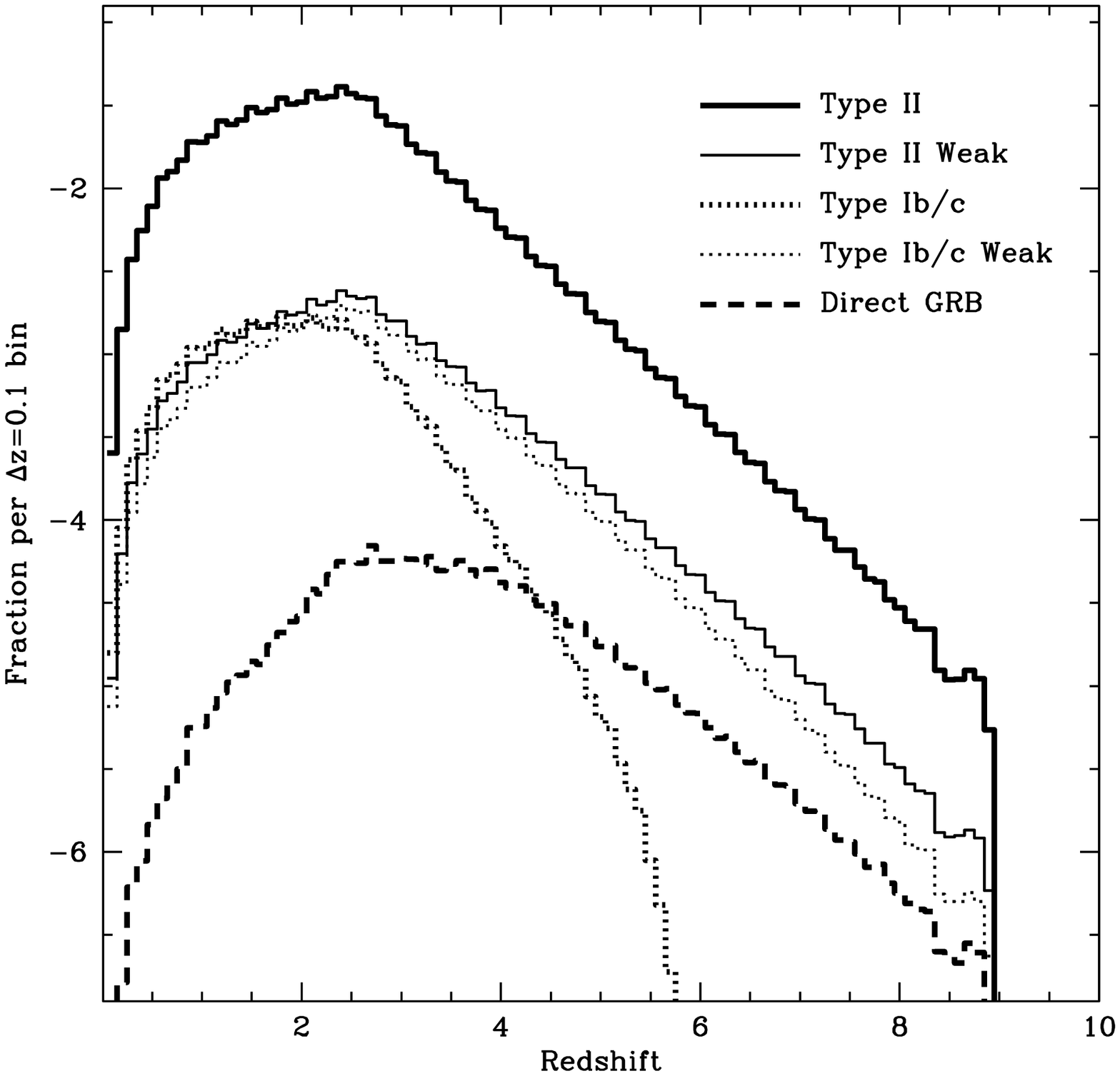}
\caption{Fraction of total outbursts per $\Delta z =0.1$ bin for 5
types of explosion: Type II supernova, Weak Type II supernova, Type
Ib/c Supernova, Weak Type Ib/c supernova, Direct-Collapse GRB.  The
weak supernovae will produce black holes.  Note that the GRB rate from
direct collapse happens at a higher redshift than core-collapse
supernovae, but the weak supernovae are not so different than the
redshift distribution of type II supernovae.  Strong Type Ib/c
supernovae have the lowest mean redshift.}
\label{fig:rate-single}
\end{figure}

For most of our studies, we will use the classic single GRB rate as
the basis for our intuition.  Figure~\ref{fig:rate-single} shows the
fraction of total outbursts per $\Delta z =0.1$ bin for 5 types of
explosion: Type II supernovae, Weak Type II supernovae, Type Ib/c
Supernovae, Weak Type Ib/c supernovae, and Direct-Collapse GRBs.  The
weak supernovae will produce black holes.  For single stars, the weak
Type Ib/c supernovae may also produce GRBs.  The binary rate is very
similar except that binaries can eject the hydrogen envelopes of the
primary star, producing more Type Ib/c supernovae, and hence, more
potential GRB systems through fallback.  Note that the peak GRB rate
from direct collapse happens at a higher redshift than core-collapse
supernovae, but the weak supernovae are not so different than the
redshift distribution of type II supernovae.  Strong Type Ib/c
supernovae have the lowest mean redshift.

We expect GRBs to be made up of both direct-collapse and fallback
black holes, but determining which type of black hole formation
scenario dominates the distribution is difficult to determine.  First
and foremost, we do not know exactly which stars have weak supernova
explosions versus the number that have no explosion~\citep{Fry99}.
Second, we know that fallback black holes will be less efficient than
direct-collapse black holes at forming GRBs.  This is because the
accretion rate in a fallback black hole system is lower (possibly
significantly lower) than a direct-collapse black hole system.  The
lower accretion rate, the weaker the power available to drive an
explosion.  \cite{Pop99} argued that the accretion rate would have to
be above 0.01$M_\odot y^{-1}$ to produce a GRB if it is
powered by neutrinos\footnote{This constraint probably also exists for
magnetically-driven outbursts at some
level.}.  Recall also that the rates for GRBs here are upper limits
(assuming all stars have enough angular momentum).

\begin{figure}
\plotone{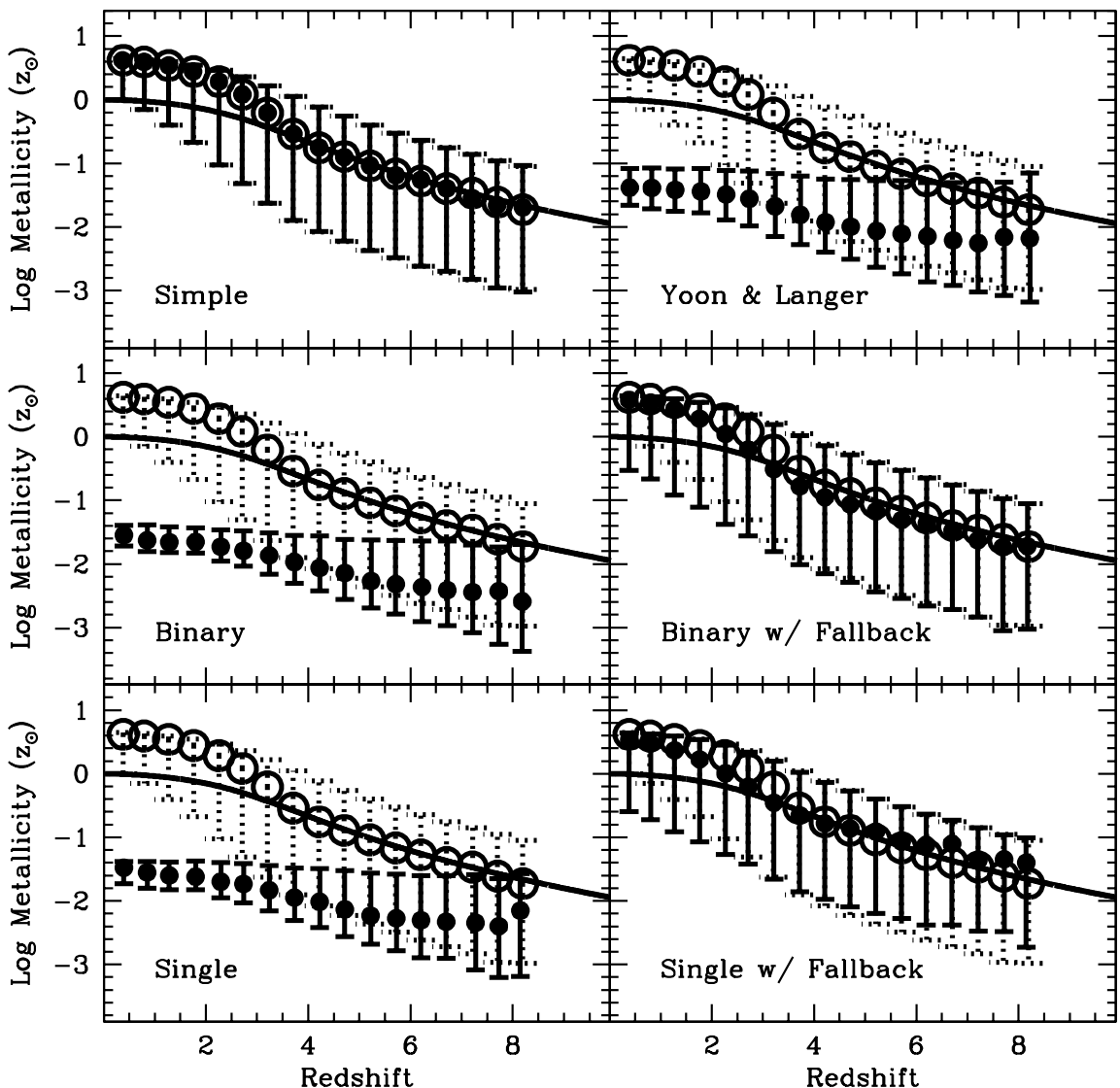} \caption{Metallicity of GRBs (solid circle and
lines) and Supernovae (open circles and dotted lines) versus redshift
for 6 separate simulations: our simple model using just a mass limit,
the~\cite{Yoo06} model prediction, Single and Binary models with and
without fallback black holes included.  The error bars correspond to
the range enclosing 90\% of all bursts.  The solid line shows the 
metallicity of an $L*$ galaxy as a reference point.  Note that without the
fallback black holes, the metallicity distribution of single and
binary stars under the~\cite{Heg03} model predicts essentially all 
bursts occur below a metallicity of 0.1, very similar to the~\cite{Yoo06} 
prediction.  But with fallback black holes, metallicities can be nearly 
as high as the supernova distribution (indeed, at high redshift, 
single stars predict metallicities higher than the supernova rate).}
\label{fig:metal6}
\end{figure}

Figure~\ref{fig:metal6} shows the metallicity of GRBs and supernovae
versus redshift for 6 separate simulations.  Note that without the
fallback black holes, the metallicity distribution of single and
binary stars under the~\cite{Heg03} model predicts essentially all
bursts occur below a metallicity of 0.1, very similar to
the~\cite{Yoo06} prediction.  This is because the~\cite{Heg03} models
argue that above this metallicity winds are sufficiently strong that
massive stars only form black holes through fallback.  With fallback
black holes, metallicities can be nearly as high as the supernova
distribution (indeed, at high redshift, single stars predict
metallicities higher than the supernova rate).  A simple explanation 
of any result that lies in between the pure direct-collapse and 
direct-collapse plus fallback systems is that the fraction of fallback 
black hole systems that form GRBs is smaller than those of direct-collapse 
black holes.  Measuring this system could tell us about how fallback 
occurs in weak supernovae.

\begin{figure}
\plottwo{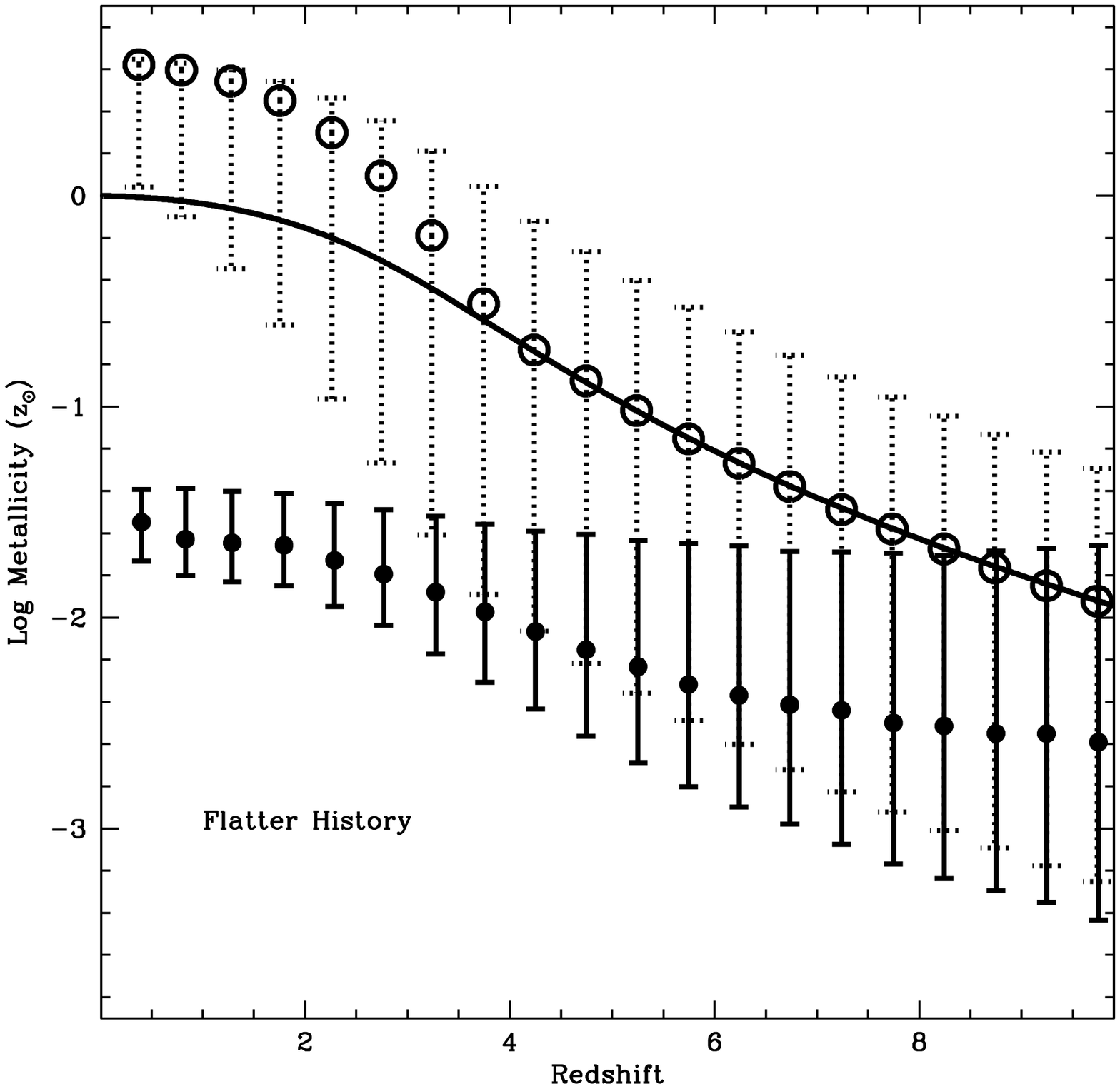}{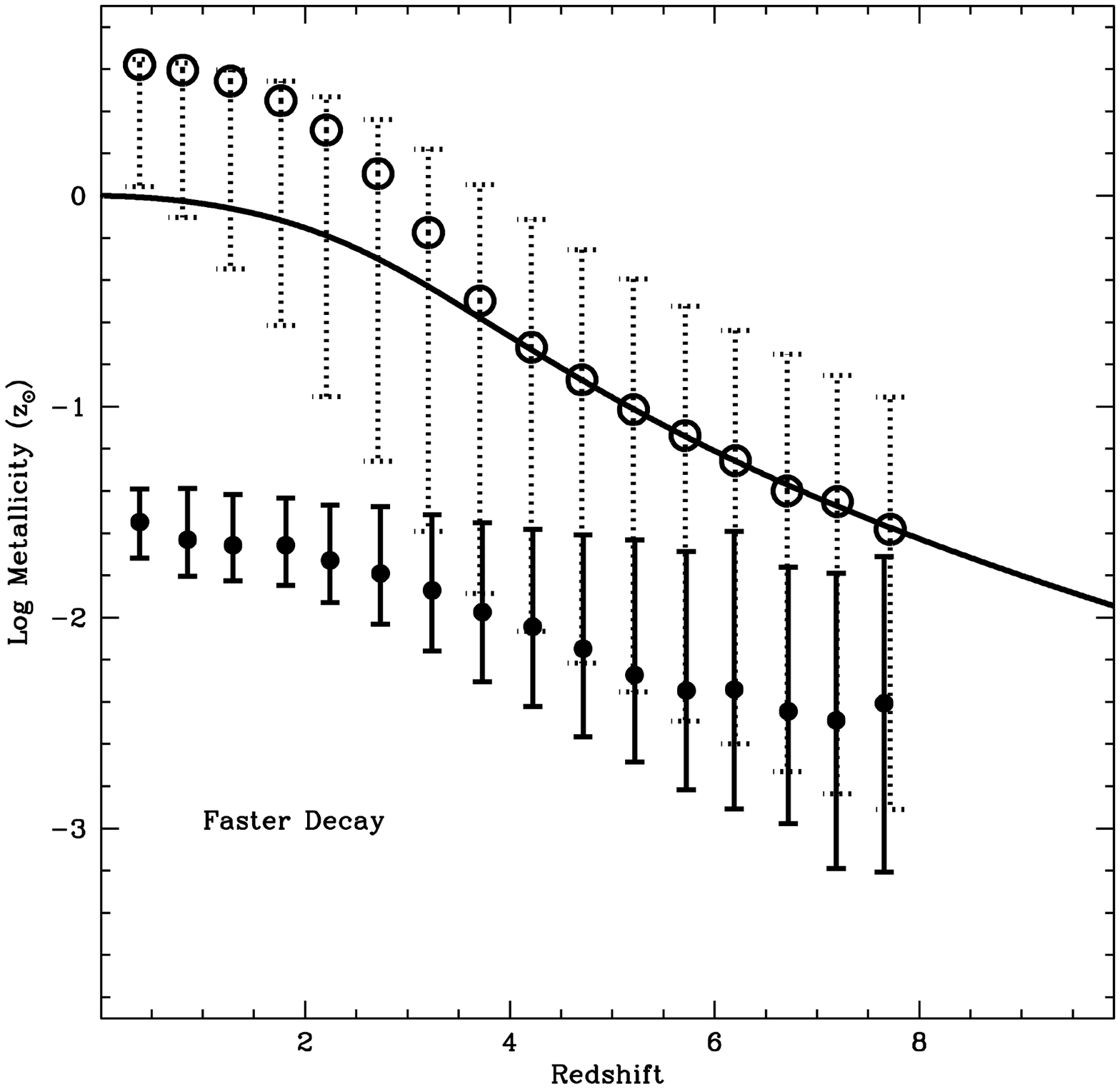}
\caption{The metallicity distribution as a function of redshift for
our direct-collapse binary GRBs and supernovae using two different
star formation histories: $A,B,z_p$ from equation~\ref{eq:sfh} are
(1.25,2.0,0.5) and (1.0,2.5,0.1) for the decay and flat histories
respectively.  Within these narrow bounds of the star formation history, 
the changes in the history are modest.}
\label{fig:binary-snr}
\end{figure}

Figure~\ref{fig:binary-snr} shows the metallicity distribution for our 
binary models for two separate star-formation histories.  Although 
the flatter star formation history allows bursts and supernovae 
to occur at higher redshift, the other changes are fairly minimal.  
Only the highest redshift bursts will really help us constrain 
the star formation history.

\begin{figure}
\plotone{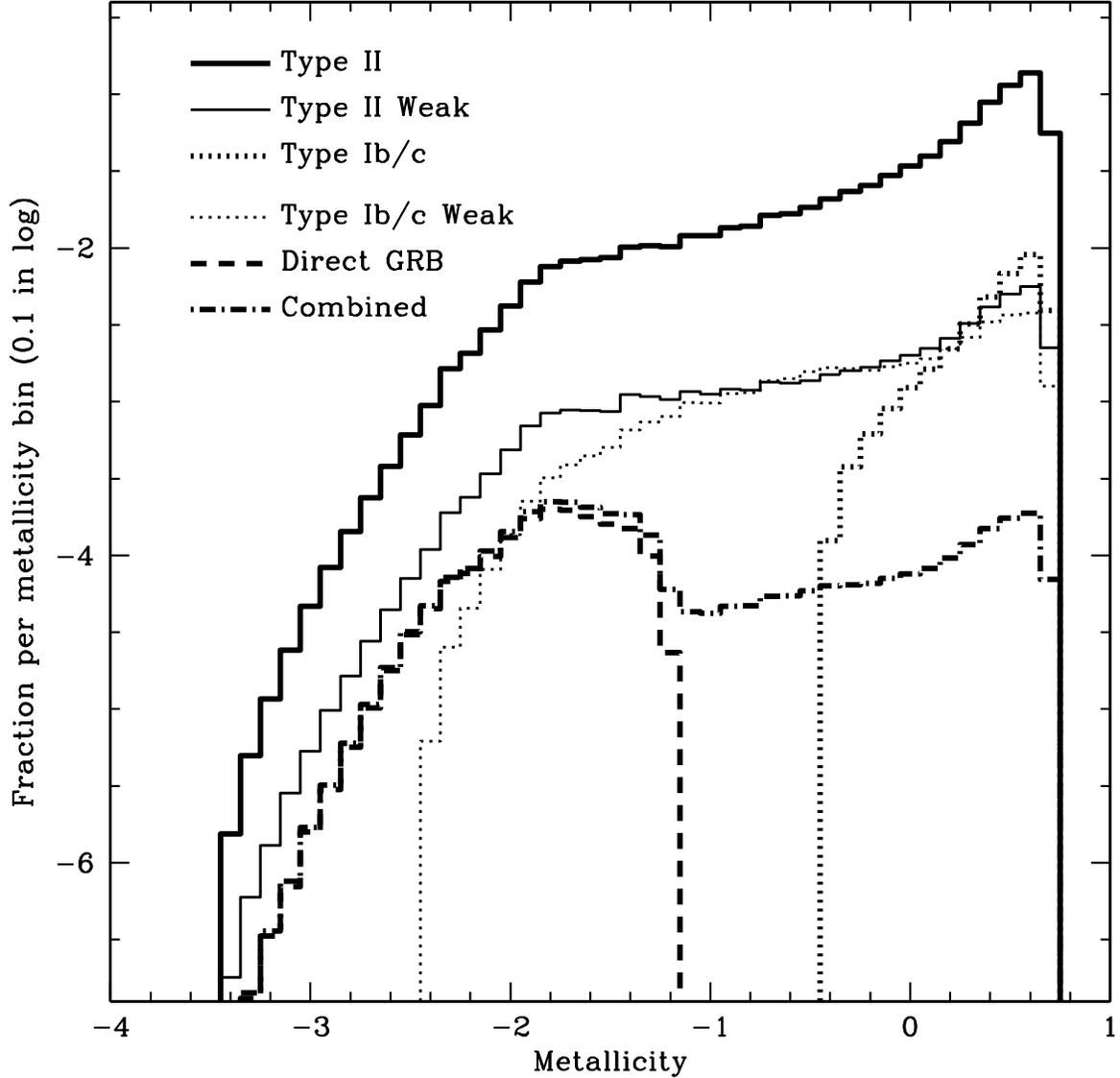}
\caption{Fraction of explosions per metallicity bin (0.1 in log space)
as a function of metallicity for single stars.  The total GRB
distribution could be the summation of the direct-collapse and weak
Type Ib/c supernovae. ``Combined'' is the sum of the direct collapse
and 2\% of the weak Type Ib/c rate. Binary systems are nearly
identical, except that their distribution could include the direct
collapse plus both types of weak supernovae.  In any event, we expect
the GRB population to have at least a slightly lower mean metallicity
than supernovae.  This mean metallicity could be significantly lower
if direct-collapse GRBs dominate the GRB rate.}
\label{fig:mevz}
\end{figure}

\begin{figure}
\plotone{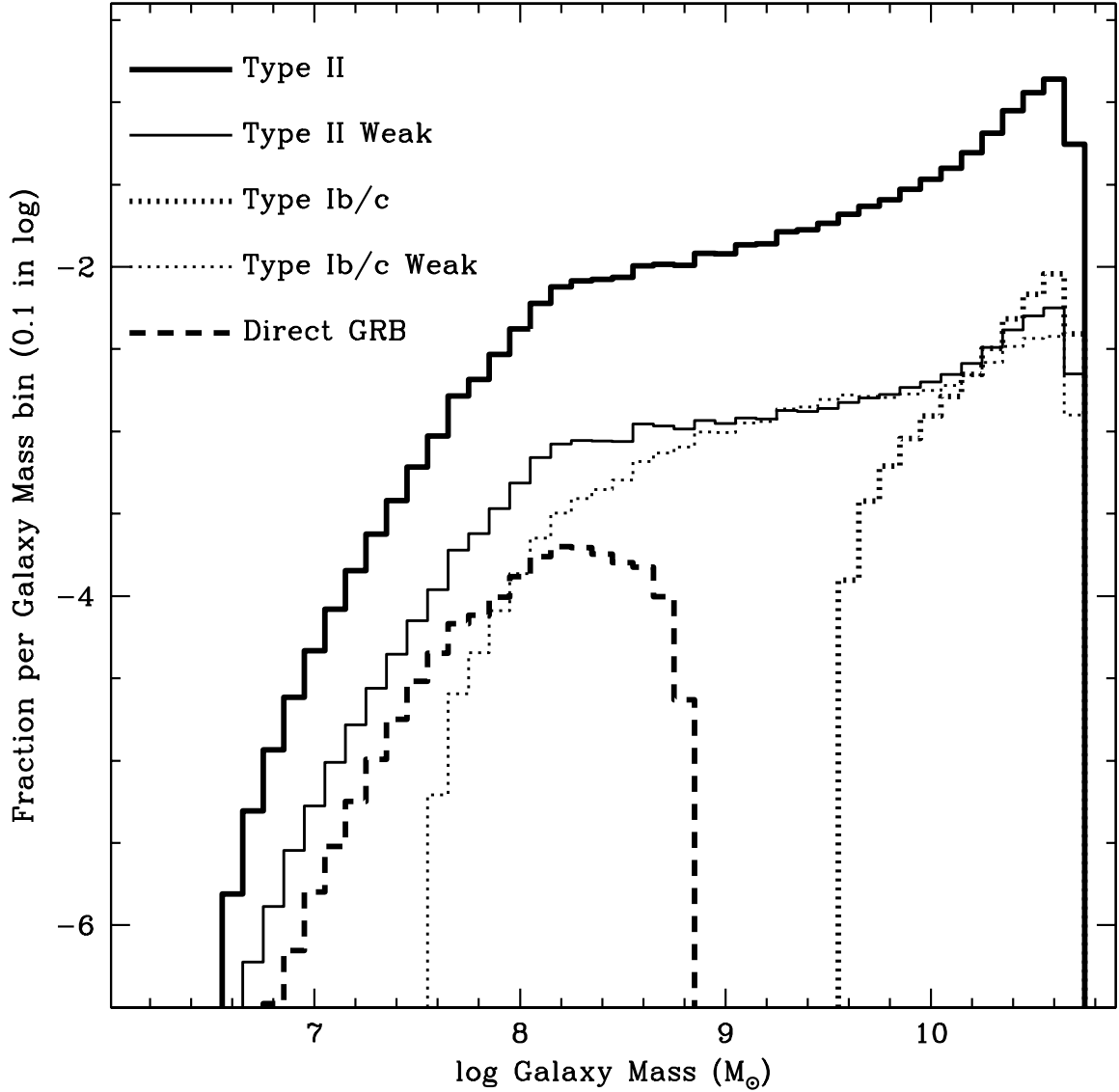}
\caption{Fraction of explosions per galaxy mass bin (0.1 in log space)
as a function of metallicity for single stars.  The total GRB
distribution could be the summation of the direct-collapse and weak
Type Ib/c supernovae.  Binary systems are nearly identical, except
that their distribution could include the direct collapse plus both
types of weak supernovae.  In any event, it is clear that it could 
well be that GRBs are produced, preferentially, in smaller mass 
galaxies.}
\label{fig:massg}
\end{figure}

Figure~\ref{fig:mevz} shows the metallicity distribution of all
explosions as a function of redshift.  The total GRB distribution
could be the summation of the direct-collapse and some fraction of weak
Type Ib/c supernovae.  Binary systems are similar, except that their
distribution could include the direct collapse plus both types of weak
supernovae.  In any event, we expect the GRB population to have at
least a slightly lower mean metallicity than supernovae.  Included in
Figure~\ref{fig:mevz} is a plot assuming showing the combined
distribution assuming 2\% of all fallback black holes produce GRBs in
the binary scenario.  This produces a mean metallicity for GRBs just a
little above 1/10th solar.  This mean metallicity could be
significantly lower if direct-collapse GRBs dominate the GRB rate.  It
is no surprise that the mean mass of GRB-producing galaxies is also
lower than normal supernovae (Fig.~\ref{fig:massg}).

Figure~\ref{fig:bmzgrb} shows the number of binary GRBs including
those from weak SN II and Ib/c(top), single star GRBs including weak
SN Ib/c (middle), and all supernovae (bottom) per $\Delta =0.1$ bin in
redshift and log metallicity ($\log (z/z_{\odot})$), normalized to the
number in the peak bin. Contours are plotted for bins with 20, 40, 60,
and 80\% of the peak number. On top
of these contours, we show the current measurements of metallicity and
redshift of GRBs associated damped Ly$\alpha$ absorbers available in
the literature \citep{pro07}. Triangles represent lower limits,
circles metallicity estimates with error bars.  The peaks of the
predicted distributions are similar, but the GRBs have a much larger extent to
low metallicities. This is greatest for the single star models. These
models require the largest progenitor masses. As expected, the
production of GRBs peaks between redshifts of 2 and 3, since that is
where our star formation rate peaks.

Figure~\ref{fig:bmzggrb} shows the number of binary GRBs including
those from weak SN II and Ib/c(top), single star GRBs including weak
SN Ib/c (middle), and all supernovae (bottom) per $\Delta =0.1$ bin in
redshift and log of host galaxy stelar mass [$\log (M/M_{\odot})$],
normalized to the number in the peak bin. Contours are plotted for
bins with 20, 40, 60, and 80\% of the peak
number. While the populations have similar peaks, a larger
fraction of GRBs occur in sub-L* galaxies. The higher specific star
formation rate in low mass galaxies produces IMFs populated all the
way up to the high mass stars that produce GRBs, while high star
formation rate massive galaxies do not populate the extreme masses as
well, producing a larger volume of relatively poorer stellar
populations. This bias also explains the bias towards low metallicity
in GRB hosts, as lower mass galaxies also have a lower metal content.

Figure~\ref{fig:combo-cont} shows contour plots similar to
Figures~\ref{fig:bmzgrb} and \ref{fig:bmzggrb} but for the
``combined'' scenario of Figure~\ref{fig:mevz}. By only including 2\% of
fallback black holes the peak of the distribution shifts to
considerably lower metallicities and host galaxy masses. This
distribution provides a reasonable approximation to the observed
metallicity distribution.

\section{Conclusions \label{conclusion}}

In this paper, we present the theoretical distribution of GRBs 
as a function of redshift and of metallicity for the basic 
GRB models.  We use current distributions of star formation, 
including the relative star formation rates of different-sized 
galaxies.  In this manner, we can directly compare the predictions 
of the basic progenitor models to that of the growing set of 
observational data.  

\begin{figure}
\includegraphics{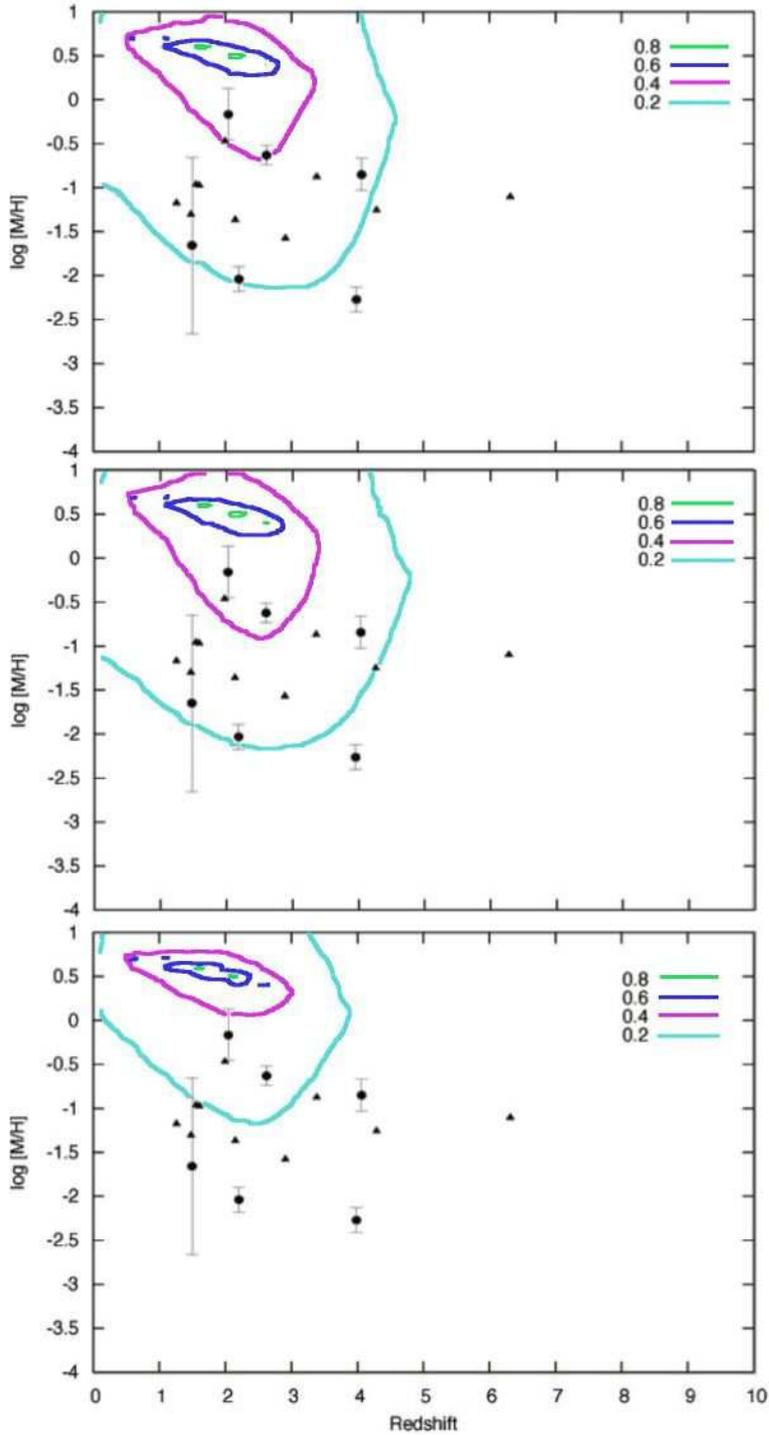}
\caption{Number of binary GRBs including those from weak SN II and
  Ib/c (top), single star GRBs including weak SN Ib/c (middle), and
  all supernovae (bottom) per $\Delta =0.1$ bin in redshift and log metallicity
  ($\log (z/z_{\odot})$), normalized to the number in the peak
  bin. Contours are plotted for bins with 20, 40, 60, and 80\% of the
  peak number.}
\label{fig:bmzgrb}
\end{figure}

\begin{figure}
\includegraphics{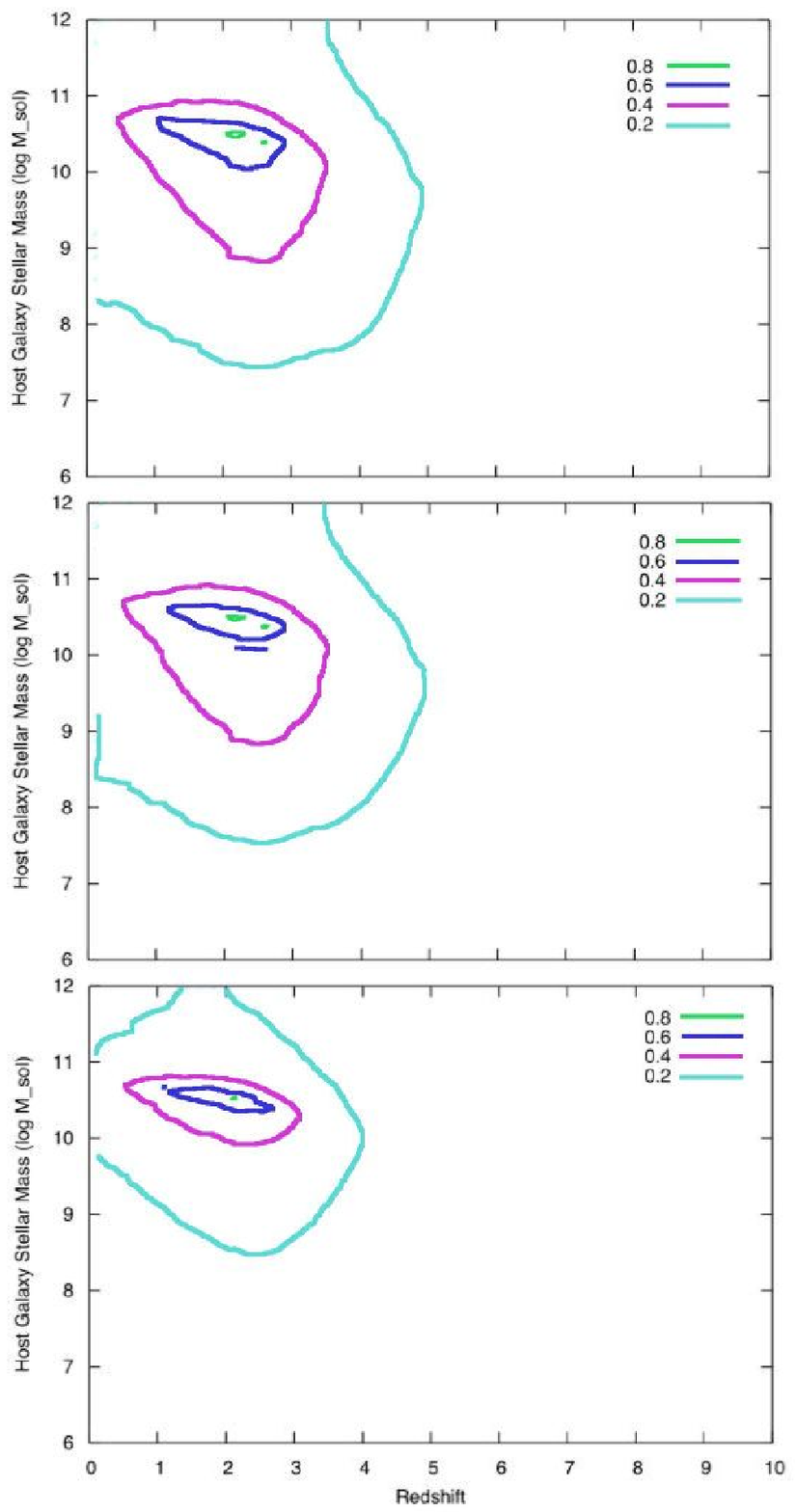}
\caption{Number of binary GRBs including those from weak SN II and
  Ib/c(top), single star GRBs including weak SN Ib/c (middle), and all
  supernovae (bottom) per $\Delta =0.1$ bin in redshift and log of
  host galaxy stelar mass ($\log (M/M_{\odot})$), normalized to the
  number in the peak bin. Contours are plotted for bins with 20, 40,
  60, and 80\% of the peak number.}
\label{fig:bmzggrb}
\end{figure}

\begin{figure}
\plottwo{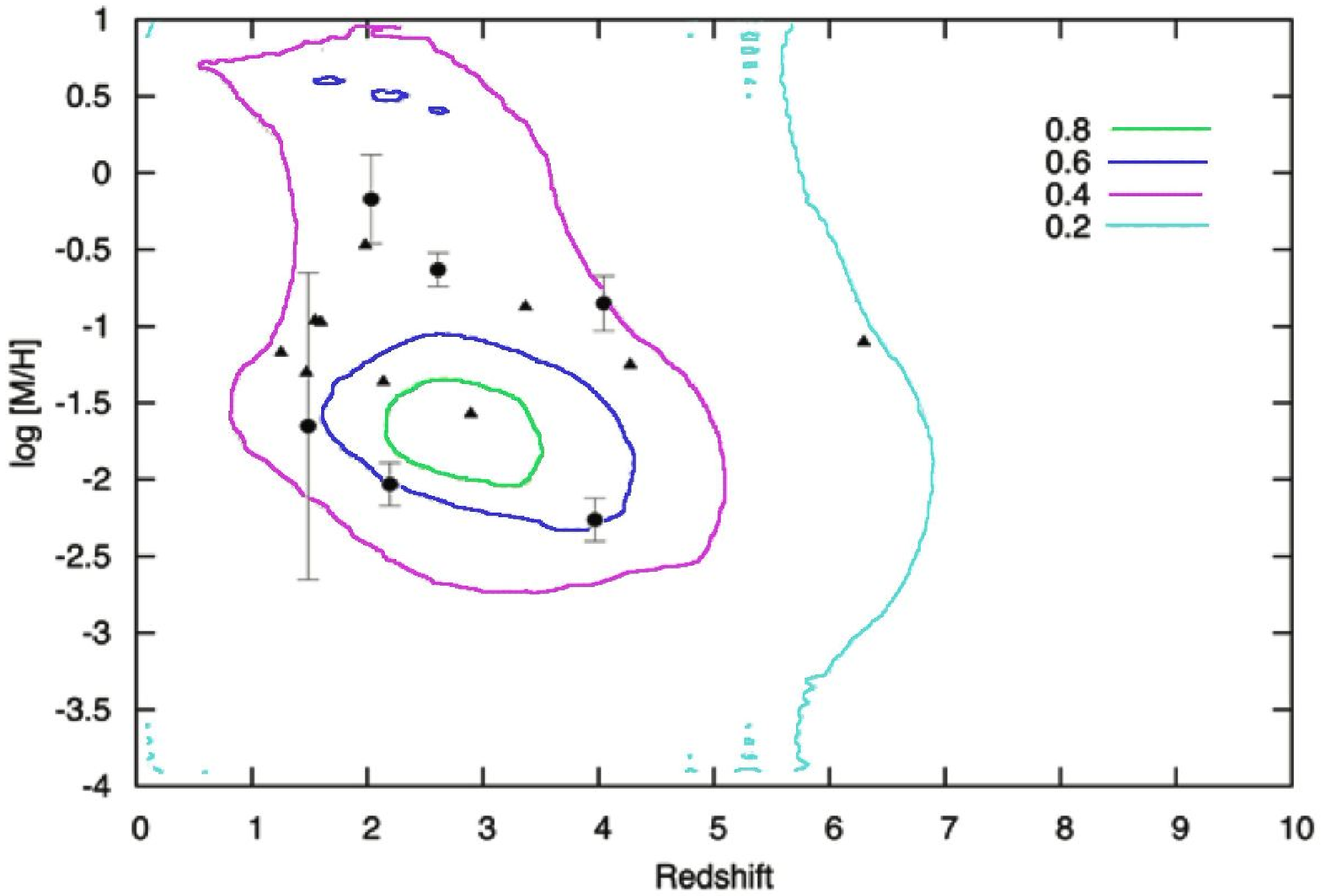}{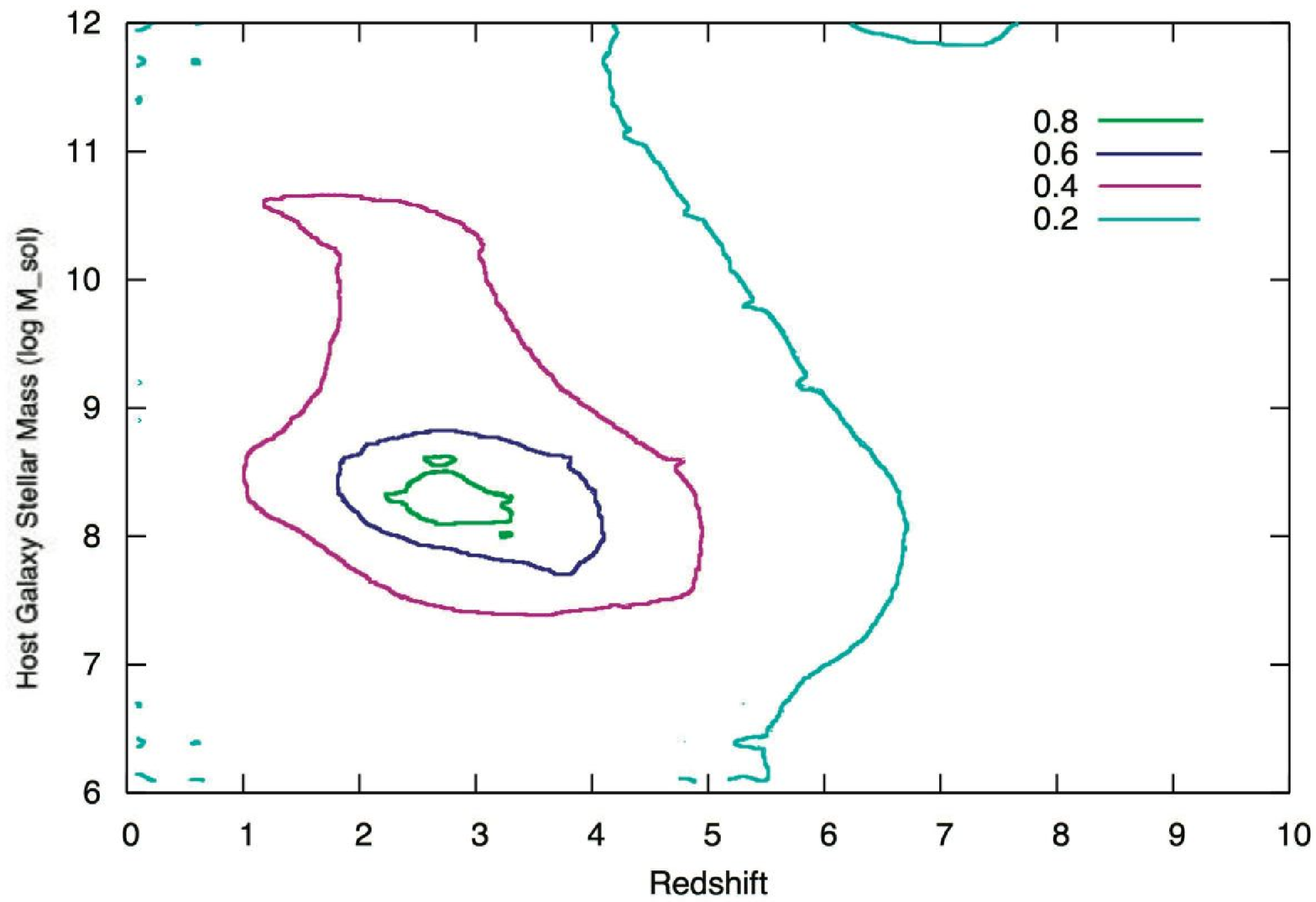}
\caption{Number of single GRBs including those from direct fallback
  and 2\% of weak SN Ib/c per $\Delta =0.1$ bin in redshift and log
  $[M/H]$ (left) and log of host galaxy stelar mass ($\log
  (M/M_{\odot})$) (right), normalized to the number in the peak
  bin. Contours are plotted for bins with 20, 40, 60, and 80\% of the
  peak number. The peaks are at much lower metallicity and host galaxy
  mass than for scenarios that include all fallback black holes. The
  observed metallicity distribution also matches better than for all
  fallback black holes or direct collapse or Yoon \& Langer models
  alone.}
\label{fig:combo-cont}
\end{figure}

\begin{figure}
\plotone{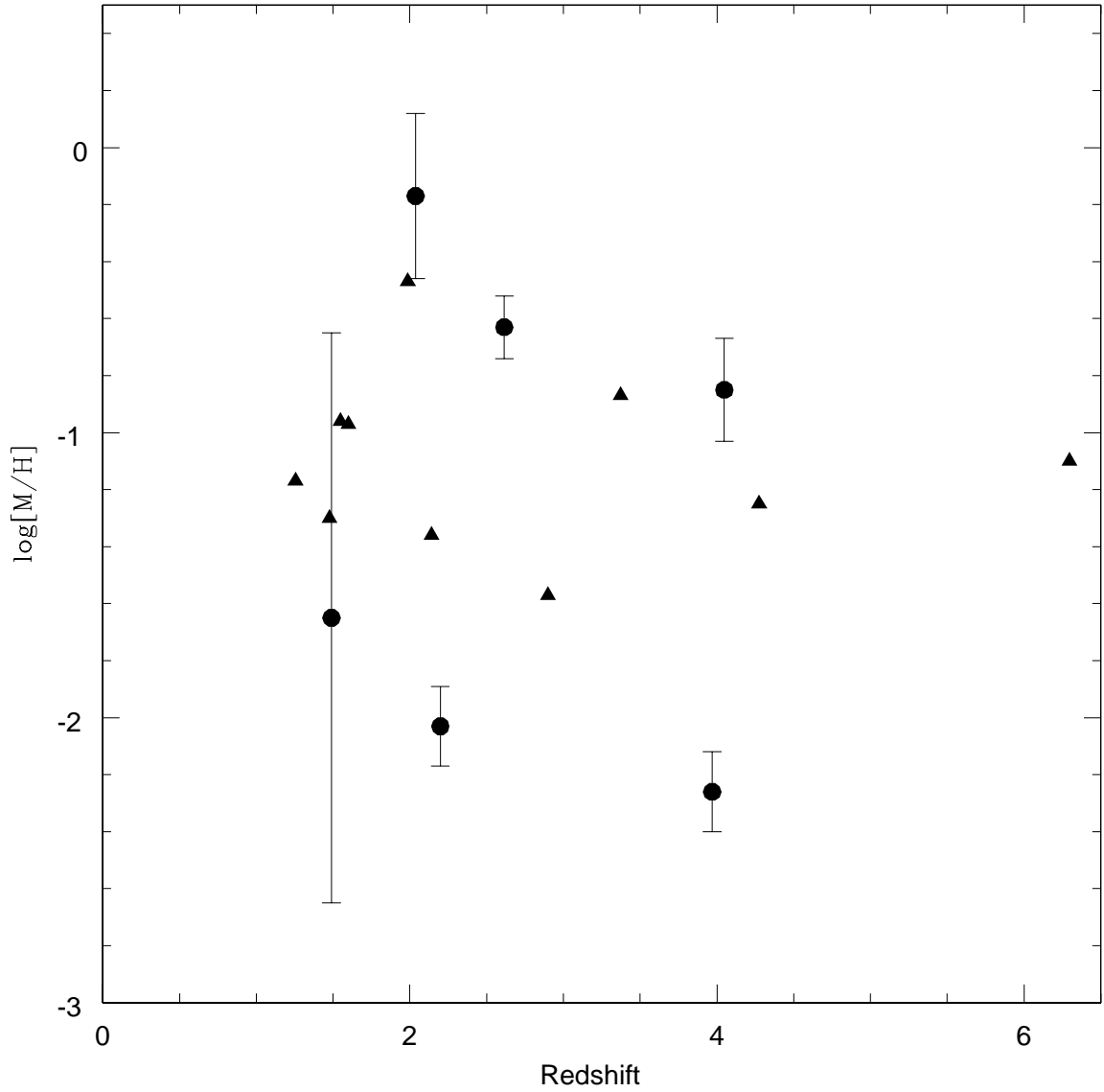}
\caption{Observed metallicity of GRB associated damped Ly$\alpha$
  absorbers versus redshift from the literature
  \citep{pro07}. Triangles represent lower limits. }
\label{fig:obsdat}
\end{figure}

From figure~\ref{fig:obsdat}, which shows the same observational data
overplotted on figure~\ref{fig:bmzgrb}, we can immediately see that
the small sample of GRBs is definitely at odds with the predictions of
our simple models.  For instance, the unmodified Yoon \& Langer
scenario cannot account for all GRBs. Their model is unable to produce
any significant number of GRBs above metallicities of 0.1 solar. If
the metallicity of GRBs without firm determinations lie even modestly
above their lower limits, then at least half the sample has
metallicities above 0.1 solar. Similarly, direct black hole scenarios
for both the binary and single star models are also biased towards low
metallicity because mass loss at high metallicity is too
extensive in the models. These models do have the advantage of producing a
relatively flat metallicity ditribution for low to moderate redshifts,
and there is no obvious trend of metallicity with redshift in the
data.

Binary and single star models that include all fallback black holes
produce too many GRBs at high metallicity by reducing the minimum mass
for producing a GRB sufficiently that the GRB distribution to a large
extent mimics the total supernova distribution.  This also can 
be ruled out by the observations.

We can make our models fit the data by assuming that the relative
fraction of GRBs from fallback black holes is roughly equal to that of
direct collapse.  This can occur either by arguing that only a small
fraction of fallback black holes have accretion rates that are high
enough to produce GRBs or by moving down the critical mass that
delineates fallback black hole formation from direct collapse black
hole formation.  Either of these possibilities could easily be true,
and unfortunately, at this time, theory has not made any definitive
predictions on either of these.

Quantitative estimates of host galaxy mass are not available for a
significant number of GRBs. However, we can examine morphologies and
see that most GRBs occur in small, actively star forming galaxies with
peculiar optical morphologies. This is consistent with a bias towards
low mass hosts. Again, the fallback scenarios produce too many GRBs at
high host galaxy mass, but a reduced fraction of fallback black holes 
could easily be made to fit the data.

There are several caveats to our current analysis.  The first is that
the current observed sample is still small and the observed
distribution may change significantly when we have a larger
statistical sample.  Until the observational biases for the data we
have can be understood and removed, it is dangerous to make any strong
conclusions.  In addition, we have made very simplistic models for the
fates of massive stars.  Binary progenitor models are much more
complex than the simple models we have used, and until we have firm
predictions from these models, it is difficult to make firm
conclusions about these progenitors.  Likewise, the most recent
single-star models will evolve with time, and perhaps the inclusion of
more complete physics will alter them to fit the data better.
Fortunately, this framework can accomodate any progenitor
prescription, so it can be used to test more sophisticated models as
they become available.

{\bf Acknowledgments} It is a pleasure to thank Aimee Hungerford and
Jason Prochaska for useful dicsussions on this project.  We also thank
Jason Prochaska for providing his latest data on GRB metallicities.
This project was funded in part under the auspices of the
U.S. Dept. of Energy, and supported by its contract W-7405-ENG-36 to
Los Alamos National Laboratory, and by a NASA grant SWIF03-0047.

{}

\newpage


\clearpage

\clearpage

\clearpage

\clearpage

\clearpage

\clearpage



\clearpage

\clearpage



\clearpage

\end{document}